\documentclass[aps,superscriptaddress,twocolumn,twoside,floatfix,pra,nofootinbib,a4paper]{revtex4}
\usepackage{times}
\usepackage{epsfig}
\usepackage{amsfonts}
\usepackage{amsmath}
\usepackage{amssymb}
\usepackage{amsthm}
\usepackage{color}
\usepackage[normalem]{ulem}
\newcommand{\stkout}[1]{\ifmmode\text{\sout{\ensuremath{#1}}}\else\sout{#1}\fi}
\usepackage{latexsym}
\usepackage{mathrsfs}
\usepackage{natbib}
\usepackage{verbatim}
\usepackage[T1]{fontenc}
\usepackage{float}

\usepackage{enumitem}

\usepackage{graphicx}
\usepackage{xcolor}

\usepackage[colorlinks=true,linkcolor=blue,citecolor=magenta,urlcolor=blue,breaklinks]{hyperref}

\newcommand{\ket}[1]{|#1\rangle}
\newcommand{\bra}[1]{\langle#1|}
\newcommand{\braket}[2]{\langle#1|#2\rangle}

\newcommand{\ketbra}[2]{|#1\rangle\langle#2|}
\newcommand{\proj}[1]{\ketbra{#1}{#1}}

\usepackage{tikz}
\usepackage{pgfplots}
\pgfplotsset{compat=1.14}

\begin{document}


\title{Compounds of symmetric informationally complete measurements and their application in quantum key distribution}


\author{Armin Tavakoli}
\affiliation{D\'epartement de Physique Appliqu\'ee, Universit\'e de Gen\`eve, CH-1211 Gen\`eve, Switzerland}

\author{Ingemar Bengtsson}
\affiliation{Stockholms Universitet, AlbaNova Fysikum 	SE-106 91 Stockholm, Sweden}

\author{Nicolas Gisin}
\affiliation{D\'epartement de Physique Appliqu\'ee, Universit\'e de Gen\`eve, CH-1211 Gen\`eve, Switzerland}

\author{Joseph M.\ Renes}
\affiliation{Institute for Theoretical Physics, ETH Z\"urich, Switzerland}

\begin{abstract}
Symmetric informationally complete measurements (SICs) are elegant, celebrated and broadly useful discrete structures in Hilbert space. We introduce a more sophisticated  discrete structure compounded by several SICs. A SIC-compound is defined to be a collection of $d^3$ vectors in $d$-dimensional Hilbert space that can be partitioned in two different ways: into $d$ SICs and into $d^2$ orthonormal bases. While a priori their existence may appear unlikely when $d>2$, we surprisingly answer it in the positive through an explicit construction for $d=4$. Remarkably this SIC-compound admits a close relation to mutually unbiased bases, as is revealed through quantum state discrimination. Going beyond fundamental considerations, we leverage these exotic properties to construct a protocol for quantum key distribution and analyze its security under general eavesdropping attacks. We show that SIC-compounds enable secure key generation in the presence of errors that are large enough to prevent the success of the generalisation of the six-state protocol. 
\end{abstract}


\maketitle


\textit{Introduction.---}
Quantum information theory has established a permanent link between the foundations of quantum theory and quantum information technologies. This has reinvigorated interest in understanding the ultimate limitations of quantum states and measurements as discrete structures in Hilbert space. Quantum states and measurements have a rich geometry that has no counterpart in classical models. Therefore, it is unsurprising that the most elegant and sophisticated discrete structures that can be found in Hilbert space frequently also are the most celebrated and useful resources for the processing of quantum information.

An outstanding example is known as a \textit{symmetric informationally complete} set of pure quantum states (SIC). A SIC is a maximal set (size $d^2$) of $d$-dimensional states, $\{\ket{\phi_k}\}_{k=1}^{d^2}$, with the property that the overlap between any pair of states has the same magnitude:
\begin{equation}\label{SIC}
	|\braket{\phi_k}{\phi_l}|^2=\frac{d\delta_{k,l}+1}{d+1},
\end{equation}
where the constant on right-hand-side is fixed by normalisation. Interestingly, a SIC can both be interpreted as a set of states (as above) and as a generalised quantum measurement (positive operator-valued measure, POVM) with $d^2$ possible outcomes. The measurement operators in such a SIC-POVM  are merely the subnormalised projectors of a SIC, namely $\{\frac{1}{d}\ketbra{\phi_k}{\phi_k}\}_{k=1}^{d^2}$.

SICs have been investigated for a long time in many different contexts \cite{Hesse, Delsarte, Hoggar, Zauner, Renes}. Their relevance in pure mathematics is remarkably diverse \cite{Galois, ApplebyFlammia, ApplebyFuchs} and they even have technological applications in high-resolution radar \cite{Radar} and speech recognition \cite{Speech}. However, their interest in physics stems from their prominent role in quantum information theory \cite{Renes}. SIC-POVMs are key tools for quantum state tomography \cite{ScottTight, Zhu, Petz}, which has motivated their experimental realisation in high-dimensional Hilbert spaces \cite{Medendorp, Pimenta, Bent}. Generally, SICs and SIC-POVMs are used in a range of protocols: quantum key distribution (QKD) \cite{Codes, Singapore, Bouchard}, entanglement detection  \cite{Kalev, Shang, Bae}, device-independent random number generation  \cite{Acin, TavakoliFarkas}, dimension witnessing \cite{Brunner} and characterisation of quantum devices \cite{TavakoliSIC, TavakoliNonProj, Massi, Piotr, TavakoliSDI}. Moreover, SICs have been studied in the context of quantum nonlocality \cite{TavakoliFarkas, Vertesi, Plato, EJM} and they have an interesting foundational role in QBism \cite{QBism}. All this has triggered much interest in addressing the existence of SICs in general Hilbert space dimensions. Presently, existence has been proven numerically at least up to $d=151$ \cite{Renes, ScottGrassl, Scott, FuchsReview} and is conjectured for any $d$ (see \cite{FuchsReview} for a review).

In this work, we introduce a natural discrete Hilbert space structure that is compounded of many separate SICs. The resulting \textit{SIC-compound} is a set of $d^3$ pure $d$-dimensional quantum states, denoted $\{\ket{\psi_{jk}}\}_{jk}$ for $j\in[d^2]$ and $k\in[d]$ (where $[s]=\{1,\ldots,s\}$) with the following two properties:
\begin{enumerate}[label=\Roman*]
	\item For every $k$, the states $\{\ket{\psi_{jk}}\}_{j}$ form a SIC. 
	\item For every $j$, the states $\{\ket{\psi_{jk}}\}_{k}$ form an orthonormal (ON) basis of Hilbert space.
\end{enumerate}
In a handy terminology, we say that a SIC-compound is composed of $d$ ``orthogonal SICs``, in the sense that elements numbered $j$ in the $d$ SICs are orthogonal to each other. Indeed, given that the existence of SICs is a longstanding open problem \cite{OpenProblems}, deciding the existence of a SIC-compound for a given $d$ is expected to be even more challenging. A priori, it may seem unlikely that SIC-compounds exist at all when $d>2$ (it turns out that $d=2$ is exceptional). We address the existence of SIC-compounds for $d=3,\ldots,8$. For $d=3$ we prove that no SIC-compound exists and for $d=5,6,7,8$ we give evidence in support of the same conclusion. Remarkably, however, for $d=4$ we are able to analytically construct a SIC-compound, thus proving that they, in fact, can exist in higher-dimensional Hilbert spaces. The many symmetries of the SIC-compound, which go beyond its defining properties, allow it to be represented as a Latin square. Moreover, we find that the SIC-compound admits a strong connection to mutually unbiased bases (MUBs) which is revealed through quantum state discrimination. Equipped with the fundamental understanding of the SIC-compound, we consider its practical application for quantum information processing. Specifically, we place the SIC-compound at the heart of protocols for QKD, analyze their security under coherent attacks and show their improved robustness as compared to the four-dimensional counterpart of the six-state protocol \cite{Bruss}  (which extends the celebrated BB84 protocol \cite{BB84}).

\textit{Qubit SIC-compound.---} It is instructive to first consider the simple example of a qubit  SIC-compound. In terms of the Bloch sphere representation, a SIC corresponds to four unit Bloch vectors such that any pair has equal magnitude overlap. Hence, the four vectors point to the vertices of a regular tetrahedron. For each vector, the unique orthogonal state is represented by the antipodal Bloch vector, and therefore the four antipodal Bloch vectors also form a regular tetrahedron. By construction, the two SICs together form a SIC-compound. Their convex hull is a cube inscribed in the Bloch sphere.

\textit{Generating SICs.---}  When $d>2$, the existence of a SIC-compound is far less clear. In order to address the matter, one  benefits much from the established knowledge of SICs which heavily exploits the Weyl-Heisenberg (WH) group. This group has two generators, $X$ and $Z$, which are required to satisfy the relations $X^d=Z^d=\openone$ and $ZX=\omega XZ $, where $\omega=e^{\frac{2\pi i}{d}}$. Every known SIC (with a single exception in dimension $8$ \cite{Hoggar}) has been obtained by applying the WH group in the following ansatz,
\begin{equation}\label{WH}
	\ket{\phi_{j}}=X^{j_1}Z^{j_2}\ket{\varphi},
\end{equation} 
for $j\equiv (j_1,j_2)\in[d]^2$ and for a suitably chosen so-called fiducial state $\ket{\varphi}$. The group generators can conveniently be chosen as the so-called shift and clock operators
\begin{align}\label{rep}
	& X=\sum_{k=0}^{d-1} \ketbra{k+1}{k} &  Z=\sum_{k=0}^{d-1} \omega^{k}\ketbra{k}{k}.
\end{align} 
For $d=2,3$ all SICs are obtained via this ansatz \cite{AllSic1, AllSic2} and the same is true for any prime $d$ provided that the SIC admits some group structure \cite{Zhuprime}. Moreover, there is numerical evidence supporting that all SICs for $d=4,5,6,7$ can be obtained via the WH group \cite{ZhuDoc}.

\textit{No qutrit SIC-compound.---} Consider the case of qutrits ($d=3$). In view of the above, by showing that no SIC-compound can be obtained via the WH group, we disprove their existence in full generality. Note that the problem is substantially simplified due to the fact that Eq~\eqref{WH} generates SICs by unitarily acting on a fiducial state. Therefore, in order to construct orthogonal SICs, we must only find orthogonal fiducial states. However, for qutrit systems there are  uncountably many relevant fiducial states \cite{Zauner, Renes} (for a fixed representation of the WH group). Fortunately, using the representation in Eq~\eqref{rep}, they all admit a simple parameterisation which allows us efficiently investigate their orthogonalities. In Appendix~\ref{Appd3}, we detail the analysis for $d=3$ and show that no more than two orthogonal SICs can be constructed. An example of two orthogonal SICs is straightforwardly obtained from choosing the two fiducial vectors $\ket{\varphi_1}=\frac{1}{\sqrt{2}}(1,1,0)^\text{T}$ and $\ket{\varphi_2}=\frac{1}{\sqrt{2}}(1,-1,0)^\text{T}$.

\textit{Ququart SIC-compound.---} For the case of ququarts ($d=4$), in contrast to qutrits, there are only $256$ fiducial states \cite{Applet} that yield SICs under the ansatz \eqref{WH} (for a fixed representation). 
Within these, one can find a SIC-compound with a simple analytical form. To present it, we change the representation of the WH-group so that the generators are written as  \cite{Cliff}
\begin{align}
&X=e^{\frac{i \pi}{4}}\begin{pmatrix}
0 & i & 0 & 0\\
-1 & 0 &0& 0\\
0&0&0&1\\
0&0&i&0
\end{pmatrix},
&Z=e^{\frac{i \pi}{4}}\begin{pmatrix}
0 & 0 & -1 & 0\\
0 & 0 &0& 1\\
i&0&0&0\\
0&i&0&0
\end{pmatrix}.
\end{align}
Note that the global phase factors only serve to ensure the correct sign of $X^d$ and $Z^d$. 
Consider also the unitary operators 
\begin{align}
&U=\begin{pmatrix}
0 & 0 & 1 & 0\\
0 & 0 &0& i\\
1&0&0&0\\
0&-i&0&0\end{pmatrix},
&V=\begin{pmatrix}
0 & 1 & 0 & 0\\
1 & 0 & 0 & 0\\
0 & 0 & 0 & -i\\
0 & 0 & i & 0
\end{pmatrix},
\end{align}
which generate a projective representation of the Klein four-group $\mathbb Z_2\times \mathbb Z_2$.  
Application of $\openone$, $U$, $V$, and $UV$ on the vector $\ket{\varphi_1}=(t,i,i,i)^\text{T}/n$ produces an orthonormal basis, where $t=\sqrt{2+\sqrt{5}}$ and $n=\sqrt{5+\sqrt{5}}$. 
Call these states $\{\ket{\varphi_k}\}_{k\in [4]}$. 
Then it can be easily verified that the states $\ket{\psi_{jk}}=X^{j_1}Z^{j_2}\ket{\varphi_k}$ form a SIC for each value of $k$, where $j=(j_1,j_2)$.
By construction, the states $\{\ket{\psi_{jk}}\}_{k\in [4]}$ form an ON-basis for each of the 16 values of $j$. 
We remark that if the computational basis is chosen as separable, all $64$ states are iso-entangled \cite{Regrouping}; the entanglement negativity is $\frac{1}{n^2}\sqrt{1+t^2}$. This constitutes an interesting parallel to the concept of iso-entangled MUBs \cite{Karol} (which upholds the same degree of entanglement per state as the SIC-compound \cite{Comment3}).

By definition, the ququart SIC-compound contains four SICs and 16 ON-bases of $\mathbb{C}^4$. Interestingly, it turns out that it upholds two additional symmetries (that have no counterpart in the qubit SIC-compound). Firstly, a careful examination of $\{\ket{\psi_{jk}}\}_{j,k}$ shows that every state is not a member of precisely one ON-basis, but in fact of two different ON-bases. Therefore, the SIC-compound houses an additional 16 ON-bases. Secondly, one finds that every state $\ket{\psi_{jk}}$ upholds the defining (SIC-like) overlap property \eqref{SIC} with $27$ other states in the SIC-compound, instead of the expected $15$. The additional $12$ SIC-like overlaps originate from an additional SIC which shares four states with the defining SIC in the compound. Thus, every state is a member of two distinct SICs (see Ref~\cite{Regrouping} and Appendix~\ref{AppMUB}) that have four elements in common.

Since we are now faced with a total of $8$ SICs and $32$ ON-bases present in the compound, one benefits from nicely organising the elements. A useful observation is that for each of the four defining SICs, one can find four sets of four states such that each is an orbit under the WH subgroup $\{\openone, X^2, Z^2, X^2Z^2\}$ (again a projective Klein four-group). By suitably permuting 
the label $j\in[16]$ in $\{\ket{\psi_{jk}}\}_{jk}$, so that $j_1$ indexes the subgroup and $j_2$ indexes the application of $\openone$, $X$, $Z$, and $XZ$, we can group these orbits together and represent the SIC-compound as a Latin square (see Figure~\ref{Figsquare}).

\begin{figure}[t]
\begin{tikzpicture}
\fill[blue!10] (0,0) rectangle (4.4,4.4);
\draw[thick,scale=1.1] (0, 0) grid (4, 4);
\foreach \x in {0,1,2,3} {
	\node[anchor=center] at (0.55+1.1*\x,0.55+1.1*\x) {\Large 4};
	\node[anchor=center] at (0.55+1.1*\x,-0.55-1.1*\x+4.4) {\Large 1};
}
\node[anchor=center] at (0.55,0.55+1.1) {\Large 3};
\node[anchor=center] at (0.55,0.55+2.2) {\Large 2};
\node[anchor=center] at (0.55+1.1,0.55) {\Large 3};
\node[anchor=center] at (2.2+0.55,0.55) {\Large 2};
\node[anchor=center] at (1.1+0.55,3.3+0.55) {\Large 2};
\node[anchor=center] at (2.2+0.55,3.3+0.55) {\Large 3};
\node[anchor=center] at (3.3+0.55,2.2+0.55) {\Large 3};
\node[anchor=center] at (3.3+0.55,1.1+0.55) {\Large 2};


\node at (2.2,4.7) {\large $k$};
\node[rotate=90] at (-0.4,2.2) {\large $j_1$ \hspace{2mm} or  \hspace{2mm} $j_2$};
\end{tikzpicture}

\caption{Schematic of the 64 states in the ququart SIC-compound. First, let us index the columns by $k\in[4]$ and the rows by $j_2\in[4]$ and let each block contain the four states $\{\ket{\psi_{jk}}\}_{j_1=1}^{4}$ (recall $j=(j_1,j_2)$). Then, each column corresponds to one of the defining SICs of the compound. The collection of elements in the identically labelled  (`1',`2',`3' and `4') blocks constitute the four additional SICs present in the compound. Secondly, let us view the Latin square as an illustration of the $16$ individual states in each row of the previous interpretation. The block with coordinates $(j_2,k)$ corresponds to the state $\ket{\psi_{jk}}$ (for any chosen row index $j_1$). Each row (of four states) then corresponds to a defining ON-basis. For $j_1=1,2,3,4$, the collection of elements with identical labels  (`1',`2',`3' and `4') constitute the total of 16 additional ON-bases present in the compound.}\label{Figsquare}
\end{figure}

\textit{On existence in $d=5,6,7,8$.---} For dimensions $d=5,6,7,8$, using the representation \eqref{rep}, there are only finitely many relevant fiducial states to be considered \cite{Comment, Applet}. The number of states that yield SICs when the WH group, in the representation \eqref{rep}, is applied to them can be regarded as known if we combine the high quality numerical results of Ref~\cite{ScottGrassl} with the group theoretical analysis of Ref~\cite{Applet}. We have enumerated all of them and exhaustively checked the number of orthogonal SICs that can be constructed using these states. We find that the number of orthogonal SICs varies ($2$, $4$, $2$ and $5$ respectively) and that no SIC-compound can be constructed. Reminding ourselves of the strong numerical evidence in support of there not existing any other SICs than those that we have explicitly constructed for $d=5,6,7$, our results render the existence of a SIC-compound for $d=5,6,7$ very unlikely. However, as previously mentioned, dimension $8$ also houses SICs that are not based on the WH group \cite{Hoggar, Anna, Stacey}. Whether a SIC-compound can be formed from these exceptional SICs is left as an open question.

Furthermore, in Appendix~\ref{AppSelf} we present a method for certifying \cite{TavakoliPrepMeas} a SIC-compound (if it exists) or falsifying their existence (if it does not exist) under the sole assumption of dimension $d$.

\textit{Discriminating the SIC-compound with MUBs.---}
The ququart SIC-compound admits a simple operational relation to a set of four MUBs. Consider that for fixed $j_2$ and $k$, we try to discriminate between the four (equiprobable) states $\{\ket{\psi_{j,k}}\}_{j_1}$. Since these states are linearly independent, we can use the ``pretty good measurement'' \cite{PGM} which is the ON-basis obtained from $\ket{\xi_{j,k}}=T^{-1/2}_{j_2,k}\ket{\psi_{j,k}}$ by varying $j_1$, where $T_{j_2,k}=\sum_{j_1}\proj{\psi_{j,k}}$. 
Measuring in this basis is in fact optimal for minimising the error probability of the discrimination, which follows from \cite{Barnett2001}.
Moreover, the resulting bases for given $j_2$ but different $k$ are identical, while the bases for different $j_2$ are mutually unbiased. Thus, the four rows of the Latin square correspond to four MUBs which, interestingly, are iso-entangled with the largest possible entanglement negativity (each basis element has an entanglement negativity of $\frac{1}{\sqrt{8}}$). In Appendix~\ref{AppMUB}, we show that the relation between the SIC-compound and the four MUBs is not a coincidence but traces back to the fact that the Clifford group contains a copy of the bipartite WH group. Finally, we note that also the fifth MUB (the computational basis) emerges from  state discrimination in the SIC-compound: a state is randomly sampled from a given column of the Latin square and we are asked to determine which row it belongs to. The optimal measurement is the computational basis.

\textit{Application in QKD.---} Let us now consider the usefulness of the $d=4$ SIC-compound in QKD. Consider a prepare and measure QKD scheme in which Alice transmits a random state $\ket{\psi_{jk}}$ and Bob randomly measures in one of the 16 defining ON bases of the SIC-compound. 
A variety of specific QKD protocols can be constructed from this starting point, depending on how Alice and Bob transform or ``sift'' their resulting data into the ``raw key''. 
Here we focus on just two sifting protocols. As in the original BB84 protocol, Alice and Bob can use the bases in the compound, taking their $k$ values as the sifted key when their $j_1$ and $j_2$ values both match. Call this Sifting B. Another possibility, which we denote Sifting A, is that $j_1$ is taken as the sifted key value when their $j_2$ values agree, but their $k$ values \emph{disagree}. (This turns out to be slightly more favorable than when the $k$ values match.)
Both protocols finish with the standard steps of parameter estimation, information reconciliation, and privacy amplification to output a secure key. 
Since both protocols use the same prepare and measure setup but differ only in the classical postprocessing, we will see that Alice and Bob can first perform parameter estimation on their data and then decide which sifting strategy to employ. 

We establish the security of both protocols against arbitrary attacks by adapting the methods of \cite{kgr,rgk,rg} to ensure security against collective attacks and then invoking \cite{post} to ensure security against arbitrary attacks. The analysis proceeds in the entanglement-based scenario of the protocol. Here Eve supplies Alice and Bob with many copies of an arbitrary bipartite state $\rho_{AB}$, to which she retains the purification in system $E$, and Alice and Bob each randomly measure the bases associated with the compound on their respective subsystems. The resulting statistics of their classical measurement choices $j$ and results $k$, as well as the possible collective attacks, are precisely the same as the prepare-and-measure scenario. 

Crucially, the symmetries of the SIC-compound translate into symmetries of both sifting protocols, and this simplifies the form of $\rho_{AB}$. As we show in Appendix~\ref{sec:qkd}, for both Sifting A and B we can assume without loss of generality that ${\rho}_{AB}=(1{-}p{-}q)\Phi_{AB}+q\pi_{AB}+p\kappa_{AB}$ for some positive parameters $q,p$ with $q+p\leq 1$, where $\Phi_{AB}$ is the maximally-entangled state, $\pi_{AB}$ is the maximally-mixed state, and $\kappa_{AB}$ is the diagonal state of perfect uniform correlation. 
In other words, the joint state is a partially depolarized and dephased maximally-entangled state.

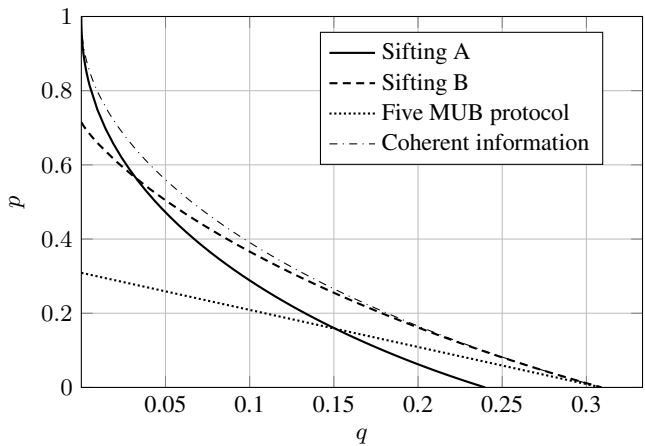
\begin{figure}[t]
\centering
\begin{tikzpicture}
\begin{axis}[
grid,
enlargelimits = false,
xlabel={$q$},
ylabel={$p$},
xmax=0.333333,
xtick={0.05,0.1,0.15,0.2,0.25,0.3},
xticklabels={0.05,0.1,0.15,0.2,0.25,0.3},
ylabel near ticks,
xlabel near ticks,
width=0.5\textwidth,
legend style={font=\small,at={(.93,0.6)},anchor=south east},
legend cell align=left,
y post scale=0.8
]
\addplot [thick] table {rnc.dat};
\addplot [thick,densely dashed] table {cbases.dat};
\addplot [thick, densely dotted] table {mubases.dat};
\addplot [dashdotted] table {coherentinfo.dat};


\legend{Sifting A, Sifting B, Five MUB protocol, Coherent information,}
\end{axis}
\end{tikzpicture}
\caption{\label{fig:regions}Regions of positive key rate for various protocols. For each $q$, the curves show the value of $p$ such that the key rate is zero. Sifting B outperforms the analog of the qubit six-state protocol using a full set of five MUBs. Sifting A can tolerate $p\to 1$ as $q\to 0$. Together, Sifting A and B nearly replicate the region of positive coherent information $-H(A|B)_{{\rho}}$ from the state $\rho_{AB}$.}
\end{figure}
 
Alice and Bob can determine both $p$ and $q$ in the parameter estimation phase as follows. 
It turns out that the probability of sifting success for Sifting A increases with increasing $q$, while the probability of error in the raw key depends on both $p$ and $q$. 
Therefore, before they commit to either sifting procedure, Alice and Bob can use their data to determine both parameters and only then decide which sifting procedure is more appropriate. 
Knowing the state $\rho_{AB}$, it is then a simple matter to apply known bounds on the rate of key extraction using information reconciliation and privacy amplification.

Fig.~\ref{fig:regions} depicts the values of $q$ and $p$ which lead to positive key rates. It also displays the region of positive key for the generalisation of the six-state protocol to $d=4$ (using a full set of five MUBs).
To enable a fair comparison, the latter protocol also discards sifting information~\cite{discard}. Its symmetries ensure that it treats all states delivered by Eve as depolarized maximally-entangled states, so that when the actual joint state is of the form $	\rho_{AB}$ above, it sees a depolarization rate of $1-p-q$. Therefore, the region of positive rate for the five MUBs protocol is symmetric under interchange of $p$ and $q$. 
Using the rate expression derived in \cite{lana}, we find the threshold for $p=0$ to be $q\approx 0.309$. This is also the threshold of the Sifting B protocol. 

\textit{Conclusions.---} We have introduced SIC-compounds as an elegant and sophisticated discrete structure in Hilbert space. Against initial intuition, we found that SIC-compounds can exist beyond qubit systems and explicitly constructed a four-dimensional SIC-compound. We found that it upholds many unexpected symmetries as well as an operational connection to mutually unbiased bases. Then, through our example of SIC-compounds, we illustrated that foundational understanding of discrete structures of quantum systems not only are interesting in themselves but that they also serve as new, powerful, tools for quantum information processing. We applied SIC-compounds towards quantum key distribution and showed that they can produce secure key in relevant situations in which the generalisation of the six-state protocol no longer is useful. 

Lastly, we ask whether four-dimensional SIC-compounds can be used to construct interesting entangled measurements of two (or more) four-dimensional systems; generalising the measurements of \cite{GisinEntropy, EJMfamily} .

\begin{acknowledgments}
This work was supported by the Swiss National Science Foundation via the NCCR-SwissMap.
\end{acknowledgments}

\appendix

\section{No SIC-compound for $d=3$}\label{Appd3}
We fix the representation of the WH group to  $X=\sum_{k=0}^{d-1} \ketbra{k+1}{k}$ and  $Z=\sum_{k=0}^{d-1} \omega^{k}\ketbra{k}{k}$. For this fixed representation, we prove that no SIC-compound exists for $d=3$. It is known that there are infinitely many fiducial states in $d=3$ \cite{Zauner, Renes}. They can be parameterised using a complete set of mutually unibased bases, which can be written (without normalisation) as follows:
\begin{align}\label{mubs}
	&\begin{bmatrix}
		1 & 0 & 0\\
		0 & 1 & 0\\
		0 & 0 & 1
	\end{bmatrix},
	& \begin{bmatrix}
		1 & 1 & 1 \\
		1 & \omega & \omega^2\\
		1 & \omega^2 & \omega
	\end{bmatrix},
	&& \begin{bmatrix}
		1 & \omega & \omega\\
		\omega & 1 & \omega\\
		\omega & \omega & 1
	\end{bmatrix},
	&&\begin{bmatrix}
		1 & \omega^2 & \omega^2\\
		\omega^2 & 1 & \omega^2\\
		\omega^2 & \omega^2 & 1
	\end{bmatrix}.
\end{align}
All \cite{AllSic1, AllSic2} fiducial states can be obtained via the following \cite{Hesse} procedure. Choose any one of the four bases. Then,  choose any pair of elements within the basis. Denote the first element by $\ket{e_1}$ and the second element by $\ket{e_2}$. The vector  $\ket{\phi}=\left(\ket{e_1}-e^{i\theta}\ket{e_2}\right)/\sqrt{2}$, for any $\theta\in[0,2\pi]$, is a valid fiducial state. Repeating this procedure for all twelve relevant pairs appearing in Eq.~\eqref{mubs}, one obtains the complete set of fiducial states.

The task of showing that no three fiducial states can form an ON-basis is significantly simplified by the fact that the problem is invariant in such a way that we can without loss of generality choose the first fiducial vector correspoding to the two first elements of the first basis in Eq.~\eqref{mubs}, namely $\ket{\phi_1}=\left(\ket{0}-e^{i\theta_1}\ket{1}\right)/\sqrt{2}$. Moreover, since every basis in Eq.~\eqref{mubs} can be transformed into every other basis in Eq.~\eqref{mubs}, it is sufficient to search for an ON-basis with respect to all fiducial states associated to, for instance, the second basis. We name the three elements of the second basis (represented in Eq.~\eqref{mubs} by the Fourier matrix) $\{\ket{f_1},\ket{f_2},\ket{f_3}\}$. Writing $\ket{\phi_2}=\left(\ket{f_1}-e^{i\theta_2}\ket{f_2}\right)/\sqrt{2}$, we straightforwardly obtain that
\begin{equation}\label{ortho}
	0\stackrel{!}{=}\braket{\phi_1}{\phi_2} \Leftrightarrow \begin{cases}
		\cos\theta_1-\cos(\theta_1+\theta_2)-\cos(\theta_2+\frac{\pi}{3})=1\\
		\sin\theta_1-\sin(\theta_1+\theta_2)-\sin(\theta_2+\frac{\pi}{3})=0.
	\end{cases}
\end{equation}
The solutions are found at $(\theta_1,\theta_2)=(2\pi/3,\pi/3)$ and $(\theta_1,\theta_2)=(5\pi/3,4\pi/3)$. To show that no third orthogonal fiducial state exists, we also consider the cases of $\ket{\phi_2'}=\left(\ket{f_1}-e^{i\theta_3}\ket{f_3}\right)/\sqrt{2}$ and $\ket{\phi_2{''}}=\left(\ket{f_2}-e^{i\theta_2}\ket{f_3}\right)/\sqrt{2}$. These give equations analogous to Eq.~\eqref{ortho}, each with two solutions. Inspecting these few cases, one easily finds that no orthogonalities exist among these solutions. Thus, we conclude that no qutrit SIC-compound exists. However, as is clear from the above, it is possible to construct two orthogonal qutrit SICs. The perhaps easiest example corresponds to the two orthogonal fiducial states
\begin{align}
	& \ket{\phi_1}=\frac{(1,1,0)^\text{T}}{\sqrt{2}}, & \ket{\phi_2}=\frac{(1,-1,0)^\text{T}}{\sqrt{2}}.
\end{align}

\section{Certification and falsification of SIC-compounds}\label{AppSelf}
We show that SIC-compounds can be certified in a semi-device-independent manner  \cite{TavakoliPrepMeas} (provided that they exist) and that existence can be disproved using  hierarchies of increasingly precise necessary conditions that each can be evaluated as a semidefinite program. 

Consider a prepare-and-measure scenario in which Alice has a random input $x\in[d^2]$ and Bob has an input $(y,y')$ which labels all pairs of elements in $[d^2]$. For convention, we take $y<y'$. Each measurement of Bob has binary outcomes $b\in[2]$. Alice's states are of dimension no greater than $d$. In Refs~\cite{Brunner, TavakoliSIC}, it was shown that the quantum maximum of the following functional
\begin{equation}
S'=\sum_{(y,y')} p(b=1| y,(y,y'))+p(b=2| y',(y,y'))
\end{equation}
is uniquely achieved in by Alice's states forming a SIC. Thus, it semi-device-independently certifies SIC preparations.  Moreover, one can add another (single) setting to Bob, $z\in[1]$, which has $o\in[d^2]$ possible outcomes, such that the modified functional
\begin{equation}
S=S'+\sum_{x=1}^{d^2} p(o=x|x,z=1)
\end{equation}
achieves its quantum maximum when both $S'$ and the above sum individually are maximal. The optimal quantum value obeys \cite{TavakoliSIC} 
\begin{align}\label{sicmax}
&\max_Q S\leq \frac{1}{2}\sqrt{d^5(d-1)^2(d+1)}+\binom{d^2}{2}+d,
\end{align}
which can be saturated if and only if Alice prepares a SIC (provided it exists) and the setting $z$ corresponds to the aligned SIC-POVM (obtained from Alice's sub-normalised preparations).

We will use this already known communication game for SICs as a building block to construct a communication game for SIC-compounds. Let Alice have inputs $x\in[d^2]$ and $i\in[d]$. Bob takes inputs $(y,y')$ and $j\in[d]$ and returns a binary outcome. Moreover, Bob  additionally has $d$ settings labelled $z\in[d]$ which have $d^2$ possible outcomes. We are only interested in cases in which $r\equiv i=j$. Let Alice and Bob play the above game (for SICs) $d$ times in parallel: each implementation (indexed by $r$) uses the preparations $\{(x,i=r)\}_x$ and the measurements $\{(y,y',j=r) \cup (z=r)\}_{y,y'}$. We label the score in the $r$'th game by  $S_r$. Naturally, these scores are so far independent since they each correspond to independent sets of preparations and measurements. If all $S_r$ are maximal, it thus certifies that Alice and Bob have implemented $d$ independent pairs of SIC preparations and SIC-POVMs. In order to certify a SIC-compound, we need to enforce the orthogonality of the $d$ SICs.

To that end, we add a penalty term. If Alice's preparation is $(x,i)$ and Bob implements one of his additional settings with $z\neq i$, then the outcome $o=x$ must never occur. If this holds true for every $(x,i,z\neq i)$, it is equivalent to a SIC-compound given that we already know that Alice must prepare SICs. Therefore, we choose our final correlation functional as
\begin{equation}
H=\frac{1}{d}\sum_{r=1}^d S_r - \sum_{\stackrel{x}{i\neq z}} p(o=x|(x,i),z).
\end{equation}
Using \eqref{sicmax} it  follows	 that 
\begin{align}
& \qquad \max_Q H \leq  \max_Q S, \qquad \text{and that}\\
& H= \max_Q S \Leftrightarrow \text{Alice prepares a SIC-compound}.
\end{align}
Thus, we have constructed a quantum communication game in which the optimal correlations are uniquely attained by SIC-compounds.

This has two notable consequences. Firstly, we may numerically search for SIC-compounds by attempting to maximise $H$ (which can be efficiently done through alternating convex searches).  Secondly, if one can prove that $H$ cannot attain the value \eqref{sicmax} in a quantum model, one falsifies the existence of any SIC-compound in the given dimension. To enable such a proof, one can use the hierarchy of semidefinite relaxations of the set of dimensionally restricted quantum correlations \cite{NV}. However, the computational requirements are significant due to the large number of preparations and measurements.  Nevertheless, semidefinite relaxations can be evaluated by employing the symmetrisation techniques of Ref~\cite{TavakoliSIC}. For instance, we consider the (trivial) case of deciding the existence of three orthogonal SICs for $d=2$. The existence of a SIC-compound would enable $H\approx 12.899$ while our semidefinite relaxation proves that no larger value is possible in quantum theory than $H\approx 12.728$. We could also evaluate the case of three orthogonal SICs in dimension three, but were unable to obtain a bound on $H$ smaller than that achieved by a SIC-compound (our SDP matrix is of size 3915). The falsification (which we have already shown analytically) could require a higher-level relaxation.

\section{SIC-compounds and MUBs in dimension four}\label{AppMUB}
Standard lore has it that SICs and MUBs are unrelated in four dimensions. SICs appear as orbits of the Weyl--Heisenberg group, and the SIC-compound is an orbit under a subgroup of the normalizer of the Weyl--Heisenberg group. 
MUBs on the other hand are obtained from the bipartite Heisenberg group. Since the two groups are different, one does not expect a connection between SICs and MUBs. Nevertheless we found a connection, and it is interesting to see how this arises.

To see this we first recapitulate the analysis by Zhu et al. \cite{Regrouping, ZhuUnpublished}, which shows that in this dimension the Clifford group contains two normal copies of the Weyl--Heisenberg group. The Clifford group contains the symplectic group $SL(2)$ with matrix elements chosen to be integers modulo 8. Its representation is fixed once the representation of the 
Weyl--Heisenberg group is fixed \cite{Applet}. The subgroup of $SL(2)$ that transforms a given compound to itself is generated by the order 4 symplectic matrices 
\begin{equation} G_1 = \left( \begin{array}{cc} 3 & 0 \\ 6 & 3 \end{array} \right) \ , \hspace{8mm} 
G_2 = \left( \begin{array}{cc} 5 & 2 \\ 2 & 1 \end{array} \right) \ , 
\end{equation}
together with an order 3 Zauner matrix \cite{Applet} which plays no role in this Appendix. The corresponding unitaries are denoted $U_{G_1}$ and $U_{G_2}$. The generators of the twin Weyl--Heisenberg group are then represented by \cite{Regrouping, ZhuUnpublished}.

\begin{eqnarray} \tilde{X} = e^{\frac{i\pi}{4}} U_{G_2} X Z = e^{\frac{i\pi}{4}} \left( 
\begin{array}{cccc} 0 & 0 & 0 & -1 \\ 0 & 0 & - 1 & 0 \\ 0 & i & 0 & 0 \\ -i & 0 & 0 & 0 
\end{array} \right) \ , \\ \nonumber \\  \tilde{Z} = U_{G_1}Z = e^{\frac{i\pi}{4}} \left( 
\begin{array}{cccc} 0 & 0 & -1 & 0 \\ 0 & 0 & 0 & i \\ i & 0 & 0 & 0 \\ 0 & 1 & 0 & 0 
\end{array} \right)\ . 
\end{eqnarray}
The presence of this `extra' Weyl--Heisenberg group explains why the $4\cdot 16$ vectors in the compound can be regrouped in such a way that $4 + 4$ SICs appear \cite{ZhuUnpublished}. 

But the bipartite Heisenberg group is lurking here as well. A straightforward calculation 
verifies that 

\begin{eqnarray} X^2 = \sigma_z \otimes \openone \ , \hspace{10mm}  
Z^2 = \openone \otimes \sigma_z \ , \hspace{2mm} \nonumber \\ \\ 
-iX\tilde{Z} = \sigma_y\otimes \sigma_y 
\ , \hspace{5mm}  Z\tilde{X} = \openone\otimes \sigma_x \ , \nonumber \end{eqnarray}

\noindent where $\sigma_x, \sigma_y, \sigma_z$ are the usual Pauli matrices. These local 
operators generate the bipartite Heisenberg group, they leave a given SIC compound invariant, 
and they can be used to create the MUBs mentioned in the main text. 

The usual construction of five MUBs proceeds by dividing the bipartite Heisenberg group into maximal abelian subgroups. In the main text we obtained 4 MUBs, all of them unbiased relative to the computational basis, as an orbit under the bipartite Heisenberg group. This is the Alltop construction of MUBs. The fact that this construction works in dimension 4 is already known \cite{Blanchfield}, but the relation to the Weyl-Heisenberg Clifford group is new.

\section{QKD security proof details
\label{sec:qkd}}

Following \cite{kgr,rgk,rg}, we can treat the sifting operation as a quantum operation as follows. 
Since the SIC-compound forms a single POVM, measurement can be described by the isometry $\ket{\phi}\mapsto \tfrac14 \sum_{jk}\ket{j}\ket{k}\braket{\psi_{jk}}{\phi}$, followed by usual projective meaurement of the $\ket{j}$ and $\ket{k}$ registers. 
Sifting can then be regarded as projective measurement of the appropriate registers, either $(j_2,k)$ or $(j_1,j_2)$, followed by postselection based on comparing the results using public communication.
Thus, each $(j_2,k)$ combination in Sifting A, for instance, gives rise to a Kraus operator $S_{j_2,k}$ which maps the $AB$ system to the raw keys $K_AK_B$ according to $S_{j_2,k}:\ket{\phi}_A\otimes \ket{\psi}_B\mapsto N\sum_{i,i'}\ket{i}_{K_A}\ket{i'}_{K_B} \bra{\psi_{i,j_2,k}}_A \bra{{\psi}^*_{i',j_2,k}}_B$, where $N$ is a normalization factor. (Recall that the conversion requires Bob to use the complex conjugate states $\ket{\psi^*_{jk}}$.)
The case of Sifting B is entirely similar.

In this formalism it is now easy to confirm that the sifting procedure is covariant under the automorphism $G$ of the SIC-compound, which is generated by $X$, $Z$, $U$, $V$, and one further unitary operator, $W$, which cyclically permutes the last three vector components and leaves the first fixed. Then, in the case of Sifting A, for any element $Y\in G$ and combination $(j_2,k)$, the operator $S_{j_2,k}Y\otimes {Y}^*=S_{j_2',k'}$ for some $j_2'$ and $k'$ (up to a phase), because the automorphism generators each preserve the individual rows and columns of the Latin square.  
Importantly, in both sifting procedures under consideration, the protocol discards the information besides the sifted key, e.g.\ the $(j_2,k)$ values in Sifting A and the $(j_1,j_2)$ values in Sifting B. 
Therefore we may average the input state $\rho_{AB}$ over $G$, since the protocol will effectively only see the state $\bar \rho_{AB}=\sum_{Y\in G} Y\otimes Y^* \rho_{AB} Y^\dagger \otimes Y^T$. 
Straightforward calculation shows that $\bar{\rho}_{AB}=(1{-}p{-}q)\Phi_{AB}+q\pi_{AB}+p\kappa_{AB}$ for some positive parameters $q,p$ with $q+p\leq 1$, where $\Phi_{AB}$ is the maximally-entangled state, $\pi_{AB}$ is the maximally-mixed state, and $\kappa_{AB}$ is the diagonal state of perfect uniform correlation.

The protocol proceeds to distill secret key from the raw key using information reconciliation and privacy amplification. 
Given a post-sifted state $\sigma_{K_AK_BE}$, we can appeal to the rate formula of \cite{dw}, $r\geq H(K_A|E)_\sigma-H(K_A|K_B)_\sigma$, where $H(K_A|E)_\sigma$ is the conditional entropy. 
The post-sifted state will be of the form 
$\sigma_{K_AK_BE}=\mathcal {M}(S_{1,1}\bar{\rho}_{ABE} S_{1,1}^\dagger)$, where $\mathcal M$ denotes the measurement of the $K_A$ and $K_B$ systems, each in the standard basis.  
This is a slight departure from and improvement on \cite{kgr,rgk,rg}, which for simplicity uses only the Bell-diagonal part of $S_{1,1}\bar{\rho}_{ABE} S_{1,1}^\dagger$. This lowers the key rate and is  unnecessary here as the state $\bar\rho_{ABE}$ itself is of a very simple form. 

\onecolumngrid


\begin{thebibliography}{99}
\bibitem{Comment3}
Note that the negativity, $\mathcal{N}$, is one-to-one with the purity, $T$, of a subsystem; $T=1-2\mathcal{N}^2$. The average purity in both SICs and complete sets of MUBs is determined by the 2-design property --- which explains their equal degree of entanglement.
\bibitem{discard} 
We could consider tomographic versions of all three protocols, as the SIC-compound is tomographically complete, but the resulting protocols are more complicated as they must estimate a large number of attack parameters.
\end{thebibliography}
\end{document}